\documentclass[lettersize,journal]{IEEEtran}
\usepackage{amsmath,amsfonts}
\usepackage{algorithmic}
\usepackage{algorithm}
\usepackage{array}
\usepackage[caption=false,font=normalsize,labelfont=sf,textfont=sf]{subfig}
\usepackage{textcomp}
\usepackage{stfloats}
\usepackage{url}
\usepackage{verbatim}
\usepackage{graphicx}
\usepackage{cite}

\usepackage{xcolor}
\usepackage{stfloats}

\begin{document}

\title{Dual-Polarized Massive MIMO Based on Precoding for Vehicle-To-Ground Communication in Urban Rail Transit}

\author{Zhengyuan Wu, Junhui Zhao,
Qingmiao Zhang, Ming Zhang
\thanks{This work was partly supported by National Engineering Research Center of System Technology for High-Speed Railway and Urban Rail Transit under Grant 2024YJ255.}
\thanks{Junhui Zhao and Zhengyuan Wu are with the school of Electronic and Information Engineering, Beijing Jiaotong University, Beijing 100044, China(e-mail:junhuizhao@hotmail.com;23111011@bjtu.edu.cn)}
\thanks{Qingmiao Zhang is with the School of Information Engineering, East China Jiaotong University ,Nanchang 330013, China(e-mail:zqm@ecjtu.edu.cn)}
\thanks{Ming Zhang is with National Engineering Research Center of System Technology for High-Speed Railway and Urban Rail Transit, Institute of Computing Technologies, China Academy of Railway Sciences Corporation Limited, Beijing 100081, China(e-mail:zm\_zhangming@hotmail.com)
}}

\markboth{}%
{Shell \MakeLowercase{\textit{et al.}}: A Sample Article Using IEEEtran.cls for IEEE Journals}


\maketitle

\begin{abstract}
The development of intelligent and diversified services in urban rail transit (URT) has resulted in an increasing demand for high-rate communication between vehicles and ground equipment. However, existing URT communication systems struggle to handle the massive data exchange required for vehicle-to-ground (V2G) communication. To address this issue, we propose a distributed dual-polarized MIMO architecture suitable for URT tunnel scenarios. Specifically, the channel model is based on spatial three-dimensional (3D) non-stationary geometry-based stochastic model (GBSM), which takes into account the geometric distribution of URT tunnels and the cross-polarization effects between dual-polarized antennas.
For dual-polarized MIMO systems, the polarized-aware sparse channel estimation (PASCE) method is proposed for effective channel estimation. Additionally, we derive closed-form expressions for the MMSE and MR precoding schemes. The polarized-aware dynamic interference cancellation (PADIC) algorithm is developed to eliminate interference between different polarization modes and multiple users. The simulation results demonstrate that the proposed dual-polarized precoding algorithm can withstand high cross-polarization correlation (XPC) and improve the efficiency of V2G communication to achieve high rates.
\end{abstract}


\section{Introduction}
\IEEEPARstart{A}{s} a core component of modern urban public transport, urban rail transit (URT) incorporates sophisticated communication technologies, playing a pivotal role in the urbanization process. The application of communication technology will also promote the development of URT system in the direction of digitalization and intelligence, such as realizing automatic driving, real-time surveillance, intelligent operation and maintenance of trains \cite{ref1}. However, the existing communication system of URT system is diversified, such as the communication based train control (CBTC) system, the long term evolution for metro (LTE-M) technology \cite{ref2}. At the same time, with the drastic increase of passenger traffic and the demand for tasks related to URT, such as passenger information service (PIS), train operation condition monitoring service (TCMS). These tasks can be categorized as latency-sensitive \cite{ref3} and computation-intensive \cite{ref4}. Therefore, the high-rate communication performance requirements for URT vehicle-to-ground (V2G) wireless systems have become more stringent. By employing 5G technologies, such as multiple input multiple output (MIMO) \cite{ref5}, URT systems can benefit from communication that offers high-rate.\par

LTE-M is based on the evolution of LTE technology, and its physical layer adopts orthogonal frequency division multiple (OFDM) technology, which can achieve good communication performance in quasi-static or low dynamic scenarios. Due to the diversity of URT operation scenarios, mainly including urban areas, suburbs, tunnels, viaducts and stations, most of the time in the tunnel environment. The high-speed operation of the train produces double dispersion channels for V2G communication, which can cause serious Doppler frequency shifts and cause inter-carrier interference (ICI) \cite{ref6}. In order to achieve reliable communication in the high mobility scenario, a novel 2D modulation technique called orthogonal time frequency space (OTFS) was proposed by Hadani \cite{ref7}. It effectively converts the dynamic channel into a 2D channel that approximates time-invariance, thus mitigating the fading of the double dispersion channel and enabling reliable communication in high-mobility scenarios \cite{ref8}. \par

The wireless coverage of the URT is primarily focused on ensuring signal connectivity within tunnels. By precisely defining the channel impact responses of both the line-of-sight (LoS) and non-line-of-sight (NLoS) paths in tunnel settings, it is feasible to model the small-scale fading characteristics of the channels. In URT environments such as depot and platform, directional antennas experience signal transmission latency due to angular errors during practical operation. Concurrently, varying environmental conditions result in uneven distribution of wireless signal. The necessity of massive MIMO has been elaborated upon in numerous references \cite{ref9},\cite{ref10}. Due to the limited space in the tunnel, the correlation between antennas increases, thereby reducing the spatial multiplexing gain of traditional co-polarized MIMO systems \cite{ref11}. As a result, research on the application of dual-polarized antennas in tunnels has also garnered significant attention. In \cite{ref12}, the author provides corresponding modeling guidance and analysis of channel capacity for dual-polarized MIMO channels. Specifically, the dual-polarized MIMO channel is modeled as a Rician fading channel, including LoS and NLoS channels, and the channel parameters are mathematically modeled. In order to accurately characterize the characteristics of information propagation in tunnel environments, scholars have conducted extensive channel modeling work in recent years \cite{ref13,ref14}.\par

Existing channel modeling methodologies are commonly categorized into deterministic \cite{ref13} and geometry-based stochastic models (GBSMs) \cite{ref14}. Considerable academic studies have employed GBSM methodologies to replicate channel dynamics scenarios \cite{ref15,ref16}. In \cite{ref17}, the B5GCM was proposed by capturing a variety of channel characteristics. It is capable of using multiple frequency bands and communication scenarios, including the high-speed train (HST).
In order to establish a unified channel modeling approach to characterize various communication scenarios, the authors proposed a pervasive wireless channel modeling theory in \cite{ref18}. In \cite{ref19},\cite{ref20}, and \cite{ref21}, channel models for rail transit scenarios are studied based on the GBSM theory. In \cite{ref19},  a 3D non-stationary MIMO GBSM for HST tunnel models was proposed. This model considers multipath effects in tunnel environments, including LoS paths, single-bounce (SB) paths, and double-bounce (DB) paths. A dynamic 3D wideband GBSM was proposed in \cite{ref20}. It described the characteristics of cooperative massive MIMO channels in intelligent high-speed railway communication systems. In \cite{ref21}, the authors proposed a non-stationary small-scale GBSM channel fading model for HST scenarios, considering various HST scenarios, and studying the birth and death processes of scattering clusters. In \cite{ref22}, a GBSM based on 3D twin clusters was proposed for underground wireless channels. This model analyses the generation and update mechanisms of different clusters in tunnel and pillar scenarios, supporting the mobility of vehicles with arbitrary trajectories and the configuration of multi-antenna systems. However, the aforementioned studies primarily focus on channel models without taking into account the influence of dual polarization on these channels.\par
The advantages of dual-polarized antennas, such as spatial multiplexing and robustness to polarization mismatch, have received increasing attention in MIMO systems. The characteristics of dual-polarized channels in tunnel environments were studied in \cite{ref23}, including cross-polarization discrimination (XPD), delay, and Doppler spread. In \cite{ref24}, the authors proposed a comprehensive solution for MIMO systems using dual-polarized antennas in frequency division duplex (FDD) mode, considering massive MIMO scenarios with dual-polarized base station antennas and multiple users. In \cite{ref25}, the authors studied the parameter characteristics of dual-polarized MIMO channels in single-cell scenarios, including XPD and cross-polarization correlation (XPC). At the same time, the authors also analyzed the spectral efficiency (SE) of MIMO systems using different precoding schemes. However, the literature discussed the establishment of channel models and dual-polarized channels separately, without conducting extensive research on dual-polarized channels based on the GBSM theory in tunnel environments.\par

Currently, the existing literature has studied leaky coaxial cable (LCX) channel models in tunnel environments \cite{ref16}. Based on the background and existing research gaps mentioned above, in this paper, we present a distributed dual-polarized antenna system to ensure comprehensive signal coverage for URT. The main contributions of this paper are summarized as follows:

\begin{itemize}
    \item A spatial three-dimensional (3D) non-stationary GBSM dual-polarized MIMO channel model is proposed for URT tunnel scenarios. This model considers the cross-polarization correlation (XPC) of a dual-polarized channel and the large-scale fading characteristics in tunnel environments.
    \item To ensure the communication rate of URT vehicles, a dual-polarized distributed MIMO system architecture is adopted. We propose a polarized-aware sparse channel estimation (PASCE) algorithm for channel estimation in URT V2G wireless communications to facilitate the design of the precoding scheme.
    \item The closed-form expressions of SE for the MMSE and MR precoding schemes are derived. To eliminate interference, we propose the polarized-aware dynamic interference cancellation (PADIC) algorithm. The impact of varying the XPC values and antenna numbers across different precoding schemes are examined subsequently. The theoretical values are also validated against actual measurements, demonstrating the superiority of the algorithm presented in this paper.
\end{itemize}\par

The remainder of this paper is organized as follows.
In Section II, the channel model and signal model are described. Section III introduces channel analysis and estimation in dual-polarized MIMO systems. In Section IV,
different precoding scheme designs and interference cancellation algorithms are analyzed. The simulation analysis and results are provided in Section V. Finally, Section VI concludes the paper.

\section{System model}
In this paper, we consider a dual-polarized distributed antenna system (DAS) of URT V2G, where the base stations (BS) are uniformly distributed along the trackside. Each BS consists of MEC, BBU and several RRUs, and each RRU's access point (AP) is equipped with a dual-polarized antenna unit. Each RRU connected to the BBU can collaborate with one another.
The train antenna unit (TAU) on the URT train is located at the top of the carriages, and there are $\cal L$ TAUs uniformly distributed along the carriages, each equipped with a dual-polarized antenna unit, which is used to serve the $\cal K$ users inside the train. The specific communication scenario model is shown in Fig.1. \par
\vspace{-0.5em}
\subsection{Channel Model Based on MIMO}
The specific MIMO spatial non-stationary geometry channel model is shown in Fig.2. The distributed APs are installed alongside the tracks at a distance of ${L_{D_p^T}}$ from the tracks and a height of ${H_{D_p^T}}$. Trains travel along tracks at a speed of $V$. The relative positions between the APs and the TAU can be utilized to determine the azimuth and elevation angles of the transmitter and receiver.

\begin{figure}[!t]
    \centering
    \includegraphics[width=3.2in]{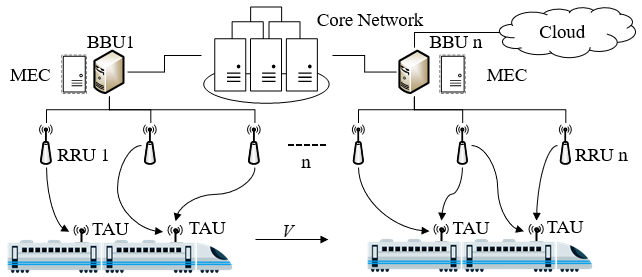}
    \caption{Distributed antenna system (DAS) for Vehicle-to-Ground (V2G) Communication in URT.}
    \label{fig1}
\end{figure}

\begin{figure*}[!t]
\centering
\includegraphics[width=0.8\textwidth]{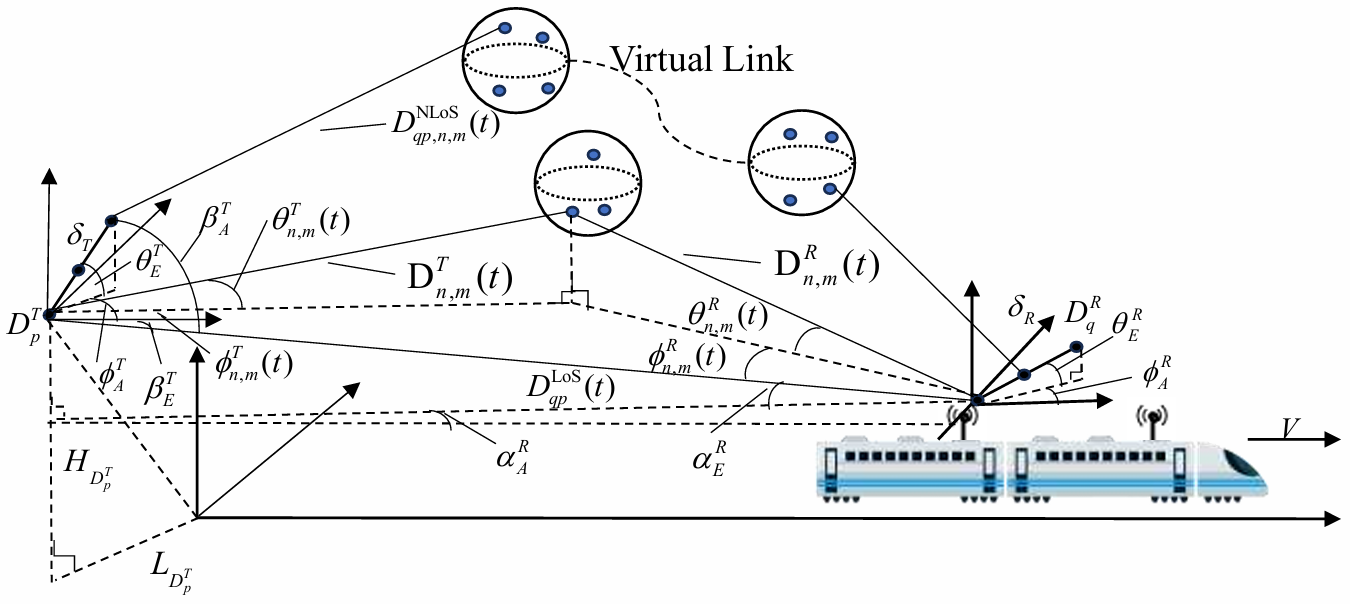}
\caption{The proposed MIMO spatial non-stationary geometry channel model for AP-TAU.}
\label{fig_2}
\end{figure*}

Based on DAS communication architecture, channel modeling is performed utilizing the GBSM method. The MIMO channel impulse response (CIR) also contains LoS and NLoS paths, which can be represented as\par
\vspace{-0.5em}
\begin{equation}
{h_{qp}}(t,\tau ) = \sqrt {\frac{K}{{1 + K}}} h_{qp}^{{\rm{LoS}}}(t,{\tau _{{\rm{LoS}}}}) + \sqrt {\frac{1}{{1 + K}}} h_{qp}^{{\rm{NLoS}}}(t,{\tau _{{\rm{NLoS}}}}),
\end{equation}
where $K$ is the Rician K-factor, the CIR of the LoS part can be expressed as
\begin{equation}
\begin{array}{l}
h_{qp}^{{\rm{LoS}}}(t,{\tau _{{\rm{LoS}}}}) = h_{qp}^{{\rm{LoS}}}(t) \cdot \delta (\tau  - \tau _{qp}^{{\rm{LoS}}}(t))\\
 = {\left[ \begin{array}{l}
{F_{q,H}}(\theta _E^R,\phi _A^R)\\
{F_{q,V}}(\theta _E^R,\phi _A^R)
\end{array} \right]^{\rm{T}}}\left[ {\begin{array}{*{20}{c}}
{{e^{j\psi _L^\theta }}}&0\\
0&{{e^{j\psi _L^\phi }}}
\end{array}} \right]\left[ \begin{array}{l}
{F_{p,H}}(\theta _E^T,\phi _A^T)\\
{F_{p,V}}(\theta _E^T,\phi _A^T)
\end{array} \right]\\
 \cdot {e^{j2\pi [{f_D}t - \tau _{qp}^{{\rm{LoS}}}(t){f_c}]}} \cdot \delta (\tau  - \tau _{qp}^{{\rm{LoS}}}(t))
\end{array},
\end{equation}
where $\theta _E^T$ and $\phi _A^T$ denote the elevation angle of departure (EAoD) and azimuth angle of departure (AAoD) of the transmitting side, respectively. $\theta _E^R$ and $\phi _A^R$ denote the elevation angle of arrival (EAoA) and azimuth angle of arrival (AAoA) of the receiving side, respectively. ${\psi _L^\theta }$, ${\psi _L^\phi }$ are random initial phase with four different polarization combinations. ${f_D}$ denotes the Doppler shift, ${f_c}$ denotes carrier frequency. The initial position vectors of the $p$th transmitting antenna element and the $q$th receiving antenna element are denoted by $D_p^T$ and $D_q^R$, respectively, which can be expressed as
\vspace{-0.5em}
\begin{equation}
D_p^T(t) = \frac{{{M_T} - 2p + 1}}{2}{\delta _T}{\left[ {\begin{array}{*{20}{l}}
{\cos (\theta _E^T)\cos (\phi _A^T)}\\
{\cos (\theta _E^T)\sin (\phi _A^T)}\\
\quad\quad{\sin (\theta _E^T)}
\end{array}} \right]^{\rm{T}}},
\end{equation}

\vspace{-0.5em}
\begin{equation}
D_q^R(t) = \frac{{{N_R} - 2q + 1}}{2}{\delta _R}{\left[ {\begin{array}{*{20}{l}}
{\cos (\theta _E^R)\cos (\phi _A^R)}\\
{\cos (\theta _E^R)\sin (\phi _A^R)}\\
\quad\quad{\sin (\theta _E^R)}
\end{array}} \right]^{\rm{T}}} + D(t).
\end{equation}

Then, the Los path between the receiver and the transmitter can be expressed as
\begin{equation}
D_{qp}^{{\rm{LoS}}}(t) = ||D_q^R(t) - D_p^T(t)||,
\end{equation}
where $|| \cdot ||$ denotes the Frobenius. $D_0$ denotes the initial distance between the transmitter and receiver, $D(t) = {D_0 + \rm{v}}_R^T(t) \cdot t$. 
\vspace{-0.5em}
\begin{equation}
{\rm{v}}_R^T(t) = v_R^T(t){\left[ {\begin{array}{*{20}{l}}
{\cos (\beta _E^T)\cos (\beta _A^T)}\\
{\cos (\beta _E^T)\sin (\beta _A^T)}\\
\quad\quad{\sin (\beta _E^T)}
\end{array}} \right]^{\rm{T}}},
\end{equation}
\vspace{-0.5em}
\begin{equation}
{{\rm{D(t)}}^T} = {D(t)^T}{\left[ {\begin{array}{*{20}{l}}
{\cos (\alpha _E^R)\cos (\alpha _A^R)}\\
{\cos (\alpha _E^R)\sin (\alpha _A^R)}\\
\quad\quad{\sin (\alpha _E^R)}
\end{array}} \right]^{\rm{T}}},
\end{equation}
${\beta _E^T}$ and ${\beta _A^T}$ denote the EAoD and AAoD of the LoS path transmitter, respectively. ${\alpha _E^R}$ and ${\alpha _A^R}$ denote the EAoA and AAoA of the LoS path receiving end, respectively. ${v_R}$ indicates the speed of the train. ${D(t)^T}$ is the ${L_2}$ norms of ${{\rm{D(t)}}^T}$.
Therefore, the delay of the LoS path can be expressed as
\vspace{-0.5em}
\begin{equation}
\tau _{qp}^{{\rm{LoS}}}(t) = \frac{{D_{qp}^{{\rm{LoS}}}(t)}}{c}.
\end{equation}\par

\begin{figure*}[hb]
	\centering
	\vspace*{-0.5em}
	\hrulefill
	\vspace*{-0.5em}
	\begin{align}
\begin{array}{*{20}{l}}
{h_{qp}^{{\rm{NLoS}}}(t,{\tau _{{\rm{NLoS}}}}) = h_{qp}^{{\rm{NLoS}}}(t) \cdot \delta (\tau  - \tau _{qp}^{{\rm{NLoS}}}(t))}\\
{ = \sqrt {\frac{{{P_{qp}}}}{\cal M}} \sum\limits_{m = 1}^{\cal M} {{{\left[ {\begin{array}{*{20}{l}}
{{F_{q,H}}(\theta _{n,m}^R,\phi _{n,m}^R)}\\
{{F_{q,V}}(\theta _{n,m}^R,\phi _{n,m}^R)}
\end{array}} \right]}^{\rm{T}}}\left[ {\begin{array}{*{20}{c}}
{{e^{j\psi _{n,m}^{VV}}}}&{\frac{{{e^{j\psi _{n,m}^{VH}}}}}{{\sqrt {{\kappa_{n,m}}} }}}\\
{\frac{{{e^{j\psi _{n,m}^{HV}}}}}{{\sqrt {{\kappa _{n,m}}} }}}&{{e^{j\psi _{n,m}^{HH}}}}
\end{array}} \right]} }\\
{}
\end{array}\left[ {\begin{array}{*{20}{l}}
{{F_{p,H}}(\theta _{n,m}^T,\phi _{n,m}^T)}\\
{{F_{p,V}}(\theta _{n,m}^T,\phi _{n,m}^T)}
\end{array}} \right] \cdot {e^{j2\pi [{f_D}t - \tau _{qp,n,m}^{{\rm{NLoS}}}(t){f_c}]}} \cdot \delta (\tau  - \tau _{qp,n,m}^{{\rm{NLoS}}}(t))
	\end{align}
\end{figure*}

Given the intricate and diverse nature of signal reflection paths, the majority of the signal energy remains focused on a select few paths. Consequently, by postulating the presence of $\cal{M}$ clusters of NLoS paths and focusing solely on the case of single-hop signal reflection, the CIR for the NLoS component can be formulated as (9), shown at the bottom of the next page. Where ${P_{qp}}$ indicates the power of the signal, ${\psi _{n,m}^{VV}}$, ${\psi _{n,m}^{VH}}$, ${\psi _{n,m}^{HV}}$, ${\psi _{n,m}^{HH}}$ are random initial phase with four different polarization combinations. ${{\kappa_{n,m}}}$ denotes the cross-polarization power ratio (XPR) between the paths of the $M$th cluster, which can be expressed as ${{\kappa_{n,m}} = {{10}^{{X_{n,m}}/10}}}$, where ${{X_{n,m}}}$ obeys Gaussian distribution with mean value $\mu $, and standard variance $\sigma $. Then, the delay of the NLoS path can be expressed as
\vspace{-0.5em}
\begin{equation}
\tau _{qp,n,m}^{{\rm{NLoS}}}(t) = \frac{{||{\rm{D}}_{n,m}^T(t)|| + ||{\rm{D}}_{n,m}^R(t)|| + D(t)}}{c} + \tilde \tau _{qp,n,m}^{{\rm{NLoS}}}(t),
\end{equation}
where ${\rm{D}}_{n,m}^T(t)$ denotes the distance of the $p$th transmitting antenna unit from the first transmitting antenna unit, 
${\rm{D}}_{n,m}^R(t)$ denotes the distance of the $q$th receiving antenna unit from the first receiving antenna unit, respectively. Both can be specifically expressed as
\vspace{-0.5em}
\begin{equation}
{\rm{D}}_{n,m}^T(t) = D_{n,m}^T(t){\left[ \begin{array}{l}
\cos \theta _{n,m}^T(t)\cos \phi _{n,m}^T(t)\\
\cos \theta _{n,m}^T(t)\sin \phi _{n,m}^T(t)\\
\quad\quad\quad{\rm{       sin}}\theta _{n,m}^T(t)
\end{array} \right]^{\rm{T}}},
\end{equation}
\vspace{-0.5em}
\begin{equation}
{\rm{D}}_{n,m}^R(t) = D_{n,m}^R(t){\left[ \begin{array}{l}
\cos \theta _{n,m}^R(t)\cos \phi _{n,m}^R(t)\\
\cos \theta _{n,m}^R(t)\sin \phi _{n,m}^R(t)\\
\quad\quad\quad{\rm{sin}}\theta _{n,m}^R(t) 
\end{array} \right]^{\rm{T}}},
\end{equation}
where $\theta _{n,m}^T(t)$ and $\phi _{n,m}^T(t)$ denote the EAoD and AAoD of the NLoS path transmitter, respectively. $\theta _{n,m}^R(t)$ and $\phi _{n,m}^R(t)$ denote the EAoA and AAoA of the NLoS path receiving end, respectively. $D_{n,m}^T(t)$ and $D_{n,m}^R(t)$ denotes the ${L_2}$ norms of ${\rm{D}}_{n,m}^T(t)$ and ${\rm{D}}_{n,m}^R(t)$, and $\tilde \tau _{qp,n,m}^{{\rm{NLoS}}}(t)$ denotes the error correction delay.

We consider the antennas of both the AP and the TAU as a uniform linear array (ULA). The spatial orientation vector of MIMO can be expressed as
\vspace{-0.5em}
\begin{equation}
\begin{array}{r}
{a_R}({\theta _{k,l}}) = [1,\exp ( - j2\pi \frac{d}{\lambda }\sin ({\theta _{k,l}})),...,\\
{\rm{               }}\exp ( - j2\pi \frac{d}{\lambda }(N - 1)\sin ({\theta _{k,l}})){]^{\mathrm{T}}}
\end{array},
\end{equation}
where $d$ denotes the antenna spacing, ${\theta _{k,l}}$ denotes the signal arrival angle.\par

Assume there are $L$ paths, ${\beta _{m,n}}(t)$ denotes the large-scale coefficient. The time-varying channel between the TAU and the AP at time $t$ can be expressed as
\vspace{-0.5em}
\begin{equation}
{\mathbf{h}_{qp}}(t) = \sum\limits_{l = 1}^L {\beta _{m,n}}(t){{h_{qp,l}}(t,\tau ){a_R}({\theta _{k,l}})a_T^H({\theta _{k,l}})} \delta (\tau  - {\tau _{qp,l}}).
\end{equation}

\subsection{Signal Model}
By considering a time-division duplex (TDD) dual-polarized massive MIMO system with $M$ subcarriers and $N$ transmitted symbols, each subcarrier having an interval of $\Delta f$, the coherence time for each symbol is $T$ \cite{ref26}. Therefore, the total bandwidth of the system is $M\Delta f$, and the duration of the symbol is $NT$. The channels remain constant within the coherent blocks and are relatively independent between different coherent blocks. Each coherent block contains ${{\tau _c}}$ symbols, where ${{\tau _p}}$ symbols are used for uplink pilot transmission to support multiuser channel estimation and the remaining symbols ${\tau _c} - {\tau _p}$ are used for data transmission. The system employs OTFS to resist double selective fading in high-speed mobile scenarios.\par

The input bits are mapped to the DD domain, forming the signals ${\mathbf{X}}_{\mathrm{DD}} \in \mathbb{C}^{M \times N}$. The ISFFT is used to perform $M$-point discrete
Fourier transform (DFT) in the delay axis, followed by $N$-point inverse discrete Fourier transform (IDFT) in the Doppler axis, facilitating the transformation into the TF domain
\vspace{-0.5em}
\begin{equation}
{\mathbf{X}}_{\mathrm{TF}} = {\mathbf{F}}_{\mathrm{M}}{\mathbf{X}}_{\mathrm{DD}}{\mathbf{F}}_N^H,
\end{equation}
where ${\mathbf{F}}_{\mathrm{M}}$ denotes the $M$-point DFT matrix with entries ${e^{ - j2\pi mn/M}}/\sqrt M$ and ${\mathbf{F}}_{\mathrm{N}}$ denotes the $N$-point IDFT matrix. After the ISFFT, the $M$-point IDFT, called Heisenberg transform, is performed along each column of the ${\mathbf{X}}_{\mathrm{TF}}$. Then the matrix in the time domain can be expressed as
\vspace{-0.5em}
\begin{equation}
{\mathbf{X}}_{\mathrm{T}} = {\mathbf{G}}_{\mathrm{tx}}{\mathbf{F}}_M^H{{\mathbf{F}}_{\mathrm{M}}}{\mathbf{X}}_{\mathrm{DD}}{\mathbf{F}}_N^H = {\mathbf{G}}_{\mathrm{tx}}{\mathbf{X}}_{\mathrm{DD}}{\mathbf{F}}_N^H.
\end{equation}\par
In this paper, a rectangular waveform ${\mathbf{G}}_{\mathrm{tx}}$ is considered, which can be expressed as ${\mathbf{G}}_{\mathrm{tx}} = diag([{g_{t}}(0),{g_{t}}(T/M), \ldots ,{g_{t}}((M - 1)T/M)])$, where ${g_{t}}(x)$ is the pulsed shaping waveform with a duration of $[0,T)$, $T$ is the sampling interval. ${\mathbf{G}}_{\mathrm{tx}}$ can be simplified to a unit matrix ${\mathbf{I}}_{M}$.\par
To mitigate inter-symbol-interference (ISI) and maintain optimal performance in OTFS communication systems, it is recommended to include a CP in each frame symbol. 
This is achieved by adding a CP matrix ${\mathbf{A}}_{\mathrm{CP}} \in \mathbb{C}^{(M + {N_{CP}}) \times M}$, where ${N_{CP}}$ represents the length of the CP. The transmitted time-domain signals can be vectorized by stacking each column, which can be expressed as $x = vec({\mathbf{A}}_{\mathrm{CP}}{\mathbf{X}}_{\mathrm{T}})$, where $x \in \mathbb{C}^{(M + {N_{CP}})N \times 1}$ is called a frame of OTFS and $vec(\cdot)$ denotes vectorization. The signal in the time domain is transmitted, and then the CP is eliminated by the CP removal matrix ${\mathbf{R}}_{\mathrm{CP}} \in \mathbb{C}^{M \times (M + {N_{CP}})}$ to obtain the OTFS symbols. The received signal is obtained in the TF domain by performing an $M$-point DFT to the individual column vectors
\vspace{-0.5em}
\begin{equation}
{\mathbf{Y}}_{\mathrm{TF}} = {\mathbf{F}}_{\mathrm{M}}{\mathbf{R}}_{\mathrm{CP}}invec(R),
\end{equation}
where $invec(\cdot)$ denote inverse vectorization. Then, the symplectic finite Fourier transform (SFFT) is performed and the DD domain signal at the receiver can be represented as follows
\vspace{-0.5em}
\begin{equation}
{\mathbf{Y}}_{\mathrm{DD}} = {\mathbf{F}}_M^H{\mathbf{Y}}_{\mathrm{TF}}{\mathbf{F}}_{\mathrm{N}}.
\end{equation}\par

We consider $L$ time-varying channels, the time-domain channel matrix $\mathbf{H}_{{\rm{OTFS}}} \in \mathbb{C}^{MN \times MN}$ can be expressed as
\vspace{-0.5em}
\begin{equation}
{{\bf{H}}_{{\rm{OTFS}}}} = \sum\limits_{l = 0}^{L - 1} {{\mathbf{h}_{qp}}(t){{\bf{\Pi }}^{{\tau _l}}}{{\bf{\Delta }}^{{\nu _l}}}} ,
\end{equation}
where ${\mathbf{h}_{qp}}(t)$, ${{\mathbf{\Pi}} _{}^{{\tau _l}}}$ and ${{\mathbf{\Delta}} _{}^{{\nu _l}}}$ indicate the CIR, the delay and Doppler shift corresponding to the $i$-th path. ${\mathbf{\Pi}} $ is the replacement matrix, which represents the effect of time delay and can be expressed as
\vspace{-0.5em}
\begin{equation}
{{\mathbf{\Pi}} _{}}=\left[\begin{array}{cccc}
{\rm{ 0}} & {\rm{ 0 }} & \cdots & {\rm{ 1}} \\
{\rm{ 1}} & {\rm{ 0 }} & \cdots & {\rm{ 0}} \\
\vdots & \ddots & \ddots & \vdots \\
{\rm{ 0}} & \cdots & {\rm{ 1}} & {\rm{ 0}}
\end{array}\right]_{NM \times NM}.
\end{equation}
 ${\mathbf{\Delta}} $ is modelling the Doppler shift, which is the diagonal matrix of $MN  \times MN$, denoted as ${\mathbf{\Delta}}  = diag[{z^0},{z^1}, \cdots ,{z^{MN - 1}}]$ with $z = {e^{j2\pi \frac{1}{{MN}}}}$.\par
The DD domain exhibits a definable input-output correlation, which can be formulated as follows
\vspace{-0.5em}
\begin{equation}
{\mathbf{Y}}_{\mathrm{DD}} = {\mathbf{H}}_{\mathrm{eff}}^{\mathrm{OTFS}}{\mathbf{X}}_{\mathrm{DD}} + \tilde {\mathbf{w}},
\end{equation}
where ${\mathbf{H}}_{\mathrm{eff}}^{\mathrm{OTFS}} = \sum\limits_{l = 1}^L {{{\mathbf{h}}_{qp}}(t)[({\mathbf{F}}_N \otimes {\mathbf{I}}_{M}){\Pi ^{{\tau _l}}}({\mathbf{F}}_N^H \otimes {\mathbf{I}}_{M})]} [({\mathbf{F}}_N \otimes {\mathbf{I}}_{M}){\Delta ^{{\nu _l}}}({\mathbf{F}}_N^H\otimes {\mathbf{I}}_{M})]$, and $\tilde {\mathbf{w}} = ({\mathbf{F}}_N^H \otimes {\mathbf{I}}_{M})w$. $\otimes$ denotes the Kronecker product. Each row of matrix ${\mathbf{H}}_{\mathrm{eff}}^{\mathrm{OTFS}}$ contains a single non-zero element, whose value and location are determined by the DD domain parameters. Moreover, there are exactly $L$ non-zero elements present in both each row and each column, i.e., it can be represented as a simple sparse structure. Represent the effective channel between the TAU and AP in the DD domain as \par
\vspace{-0.5em}
\begin{equation}
{\bf{H}}_{{\rm{eff}}}^{{\rm{OTFS}}} = \sum\limits_{l = 1}^L {{\mathbf{h}_{qp}}(t){{\bf{T}}_l}}.
\end{equation}\par
 The channel estimates obtained can be applied to precoding in the downlink, leveraging the channel reciprocity. In Section III, we will delve into the channel estimation process for dual-polarized MIMO systems.\par

\section{Dual-polarized channel analysis and estimation}
The propagation channel between the ground AP and the TAU on the train can be represented as \par
\vspace{-0.5em}
\begin{equation}
{{\rm{H}}_{k,m}} = \left[ {\begin{array}{*{20}{c}}
{h_{k,1}^{VV}}&{h_{k,1}^{VH}}& \cdots &{h_{k,{M}_{a}}^{VV}}&{h_{k,{M}_{a}}^{VH}}\\
{h_{k,1}^{HV}}&{h_{k,1}^{HH}}& \cdots &{h_{k,{M}_{a}}^{HV}}&{h_{k,{M}_{a}}^{HH}}
\end{array}} \right].
\end{equation}\par
Channel cross-polarization discrimination (XPD) refers to the channel's capacity to preserve the integrity of the transmitted or received polarization, distinguishing between horizontally (H) and vertically (V) polarized signals. Accounting for the interference due to the mutual coupling between antennas, XPD quantifies the channel's effectiveness in distinguishing between these two polarizations. XPD can be expressed specifically as  \par
\vspace{-0.5em}
\begin{equation}
{\rm{XP}}{{\rm{D}}_k}{\rm{ = }}\frac{\mathbb{E}{\left\{ {|h_{k,m}^{VV}{|^2}} \right\}}}{\mathbb{E}{\left\{ {|h_{k,m}^{HV}{|^2}} \right\}}} = \frac{\mathbb{E}{\left\{ {|h_{k,m}^{HH}{|^2}} \right\}}}{\mathbb{E}{\left\{ {|h_{k,m}^{VH}{|^2}} \right\}}}.
\end{equation}\par
The cross-polar correlation (XPC) is usually defined as the correlation coefficient between the complex gain of a vertically polarized and horizontally polarized channel, reflecting the degree of coupling of polarized signals due to multipath effects in the propagation environment, and can be expressed as follows
\vspace{-0.5em}
\begin{equation}
{\rho _{{\rm{XPC}}}} = \frac{\mathbb{E}{\left\{ {h_{k,m}^{VV}{h^ * }_{k,m}^{VH}} \right\}}}{{\sqrt {\mathbb{E}\left\{ {|h_{k,m}^{VV}{|^2}} \right\}\mathbb{E}\left\{ {|h_{k,m}^{VH}{|^2}} \right\}} }}.
\end{equation}\par

In \cite{ref25}, the polarization correlation matrix defining the correlation between channel coefficients is expressed as
\vspace{-0.5em}
\begin{equation}
{\rm{C}}_{{t_k}}^{{\rm{XPC}}} = \left[ {\begin{array}{*{20}{c}}
1&{{\rho _{{t_k}}}}\\
{\rho _{{t_k}}^ * }&1
\end{array}} \right],
{\rm{C}}_{{r_k}}^{{\rm{XPC}}} = \left[ {\begin{array}{*{20}{c}}
1&{{\rho _{{r_k}}}}\\
{\rho _{{r_k}}^ * }&1
\end{array}} \right],
\end{equation}
where ${\rho _{{t_k}}}$ and ${\rho _{{r_k}}}$ can be expressed analogously using Equ. (25). The magnitude of the XPC's strength is commonly indicated by the off-diagonal components of the channel matrix. Consequently, any alterations in elements ${{\rho _{{t_k}}}}$ and ${{\rho _{{r_k}}}}$ will have a direct impact on the channel's XPC, thus affecting the system's SE. The off-diagonal elements play a role in determining the condition number of the channel matrix, with the magnitude of this condition number serving as an indicator of the channel's health. For a 2x2 dual-polarized MIMO system, the channel matrix can be represented as follows\par
\vspace{-0.5em}
\begin{equation}
{\rm{H}}_{k,m} = \left[ {\begin{array}{*{20}{c}}
{{h_{1,1}}}&{{h_{1,2}}}\\
{{h_{2,1}}}&{{h_{2,2}}}
\end{array}} \right].
\end{equation}\par
Every element ${h}_{k,m}$ within set ${\rm{H}}_{k,m}$ comprises four distinct components of ${{\rm{H}}_{k,m}}$, identified as $h_{k,m}^{VV}$, $h_{k,m}^{VH}$, $h_{k,m}^{HV}$, and $h_{k,m}^{HH}$. The channel matrix ${\rm{H}}_{k,m}$ can be decomposed as ${\rm{H}}_{k,m} = {\rm{U}}\Sigma {{\rm{V}}^H}$, where ${\rm{U}} \in {^{{N_r} \times {N_r}}}$ and ${\rm{V}} \in {^{{N_t} \times {N_t}}}$ are unitary matrices. $\Sigma  \in {^{{N_r} \times {N_t}}}$ is a diagonal matrix with non-negative singular values for the diagonal elements, ${\sigma _1} \ge {\sigma _2}$, specifically.
Define the condition number (CN) of the channel matrix as the ratio of the maximum singular value to the minimum singular value of the matrix, which can be expressed as
\vspace{-0.5em}
\begin{equation}
\Upsilon ({\rm{H}}_{k,m}) = \frac{{{\sigma _{\max }}}}{{{\sigma _{\min }}}},
\end{equation}
where ${{\sigma _{\max }}}$ and ${{\sigma _{\min }}}$ are the maximum and minimum singular values of the matrix, respectively. In this paper, we examine the impact of varying XPC levels on SE, utilizing the variable $\Upsilon ({\rm{H}}_{k,m})$ to represent these levels. 

In summary, the propagation channel between the terrestrial BS and the users on the train can be concretely represented as follows
\vspace{-0.5em}
\begin{equation}
{{\rm{H}}_{k,m}} = \sum\nolimits_k { \odot (\sqrt {{\rm{C}}_{{r_k}}^{{\rm{XPC}}}} {\mathbf{h}_{qp}}\sqrt {{\rm{C}}_{{t_k}}^{{\rm{XPC}}}} )},
\end{equation}
where $ \odot $ denotes the product of the Hadamard product. $\sum\nolimits_k$ denotes the XPD matrix, comprising parts $h_{k,m}^{VV}$, $h_{k,m}^{VH}$, $h_{k,m}^{HV}$, and $h_{k,m}^{HH}$. Eq. (29) models the non-stationary dual-polarization MIMO channel. In this model, every channel coefficient is affected by the respective XPD, XPC, XPR, large-scale and small-scale parameters. \par

The BS at trackside needs to receive and process uplink signals, as well as appropriately precode the downlink signals. In order to estimate the channel of a single-antenna or multi-antenna system under bi-selectivity conditions, this paper employs the superimposed pilot in the DD domain of OTFS for channel estimation. A sequence of $k$ mutually orthogonal pilots ${\bf{\psi }}$ is set up to obtain the CSI using channel reciprocity. Each user in the train transmits both the data signals and the pilot signals to the BS.\par
Since the DD of OTFS exhibits sparsity, and the channel responses in different polarization directions of the dual-polarized channel also exhibit sparsity, the entire channel can be regarded as having a block sparse structure. The channel observation model can be written as
\vspace{-0.5em}
\begin{equation}
{\bf{y}} = \sum\limits_{p \in \{ V,H\} } {\sum\limits_{q \in \{ V,H\} } {{\Xi _{pq}}{{\rm{H}}_{pq}} + {\bf{w}}} },
\end{equation}
where ${{\Xi _{pq}}}$ represents a dictionary matrix, it is a block diagonal form, which can be expressed as
\vspace{-0.5em}
\begin{equation}
{{\bf{\Xi }}_{pq}} = \sqrt {{\eta _{pq}}} \left[ {{\bf{T}}_{pq}^{(1)}{\bf{\psi }}, \ldots ,{\bf{T}}_{pq}^{(L)}{\bf{\psi }}} \right],
\end{equation}
where ${{\eta _{pq}}}$ is the power control coefficient. We conceptualize channel estimation as a task of sparse signal recovery, employing dictionary matrices ${\bf{\Phi }}$ for this purpose \cite{ref27}. The sparse channel estimation model is defined as
\vspace{-0.5em}
\begin{equation}
{\mathbf{y}} = {\mathbf{\Phi}}{\mathbf{h}_s^p} + {\mathbf{w}},
\end{equation}
where ${\bf{\Phi }} = \sqrt {{\rm{R}}_{k,p}} ({{\bf{\Xi }}_{pq}} \otimes {{\bf{I}}_L})$, ${{\rm{R}}_{k,p}}$ is the spatial correlation matrix, $\mathbf{h}_s^p$ is a sparse vector consisting of four polarized channel vectors connected in series, ${\bf{h}}_s^p = {[h_{VV}^T,h_{HH}^T,h_{VH}^T,h_{HV}^T]}$.

\renewcommand{\algorithmicrequire}{\textbf{Input:}}
\renewcommand{\algorithmicensure}{\textbf{Initialize:}}
\begin{algorithm}[t]
\caption{Polarized-Aware Sparse Channel Estimation}
\begin{algorithmic}[1]
\REQUIRE Received signal $\mathbf{y}$, 
         Dictionary matrix $\mathbf{\Phi}$,
         Sparsity level $\mathbf{S}$,
         Polarization components $\mathcal{P}$.

\ENSURE $\epsilon$, $\text{max\_iter}$, 
         $\text{reduction\_factor}$, $\text{threshold}$,
         Step size = $\mathbf{S}$,
         Support set $\mathcal{P}_s = \emptyset$,
         Residual $\mathbf{y}_i^p = \mathbf{y}$,
         Polarized channel vectors $\mathbf{h}_s^{\mathcal{P}} = \mathbf{0}$.
    
    \FOR{$\text{iter} = 1$ to $\text{max\_iter}$}
        \FOR{each polarization component $p \in \mathcal{P}$}
            \STATE $\text{cor}^p = |\mathbf{\Phi}^T \cdot \mathbf{y}_i^p|$ ;
            \STATE $\text{sorted\_idx}^p = \text{sort}(\text{cor}^p, \text{descending})$;
            \STATE $\text{support}^p = \text{sorted\_idx}^p[1 : \mathbf{S}]$;
        \ENDFOR
        
        \STATE $\text{joint\_support} = \bigcup\limits_{p \in \mathcal{P}} \text{support}^p$ ;
        \STATE $\mathcal{P}_s = \mathcal{P}_s \cup \mathbf{\Phi}(:, \text{joint\_support})$;

        \STATE Calculate $\mathbf{h}_s^p$ using (33);
        
        \STATE $\mathbf{y}_i^p = \mathbf{y} - \sum\limits_{p \in \mathcal{P}} \mathbf{\Phi}^p \mathbf{h}_s^p$ ;
        \STATE $\text{residual\_reduction} = \|\mathbf{y}_i\| / \|\mathbf{y}\|$;

        \IF{$\text{residual\_reduction} \geq \text{threshold}$}
            \STATE $\mathbf{S} \gets \max(1,  \mathbf{S} \times \text{reduction\_factor})$;
            \STATE $\|\mathbf{y}_j\| = \|\mathbf{y}_i\|$;
            \IF{$\text{iter} \geq \text{max\_iter}$ \textbf{or} $\|\mathbf{y}_j\| \leq \epsilon \cdot \|\mathbf{y}\|$}
                \STATE break
            \ENDIF
        \ENDIF
    \ENDFOR
    
    \STATE \textbf{Output:} Recovered polarized channel vectors $\hat{\mathbf{h}}_s^{\mathcal{P}}$, Estimated channel matrix $\hat{\mathbf{H}}$.
\end{algorithmic}
\end{algorithm}
 
The proposed polarized-aware sparse channel estimation (PASCE) is shown in Algorithm 1. It is based on the orthogonal matching pursuit (OMP) algorithm \cite{ref10}. In solving the sparse coefficient vector, the least squares method is used to solve the optimization problem, and a regularization term is added, which serves to stabilize the solution process and avoid numerical instability. It can be represented as 
\vspace{-0.5em}
\begin{equation}
\mathbf{h}_s^p = \left( (\mathbf{\Phi}_{\mathcal{P}_s}^p)^H \mathbf{\Phi}_{\mathcal{P}_s}^p + \sigma_n^2 \mathbf{I} \right)^{-1} (\mathbf{\Phi}_{\mathcal{P}_s}^p)^H \mathbf{y}.
\end{equation}\par
The covariance of the channel estimate is ${\Gamma _{k,p}} = {P^{{\rm{ul}}}}{\tau _p}{{\rm{R}}_{k,p}}{(\sum\nolimits_{k = 1}^{\cal K} {{P^{{\rm{ul}}}}{\tau _p}{{\rm{R}}_{k,p}} + {\sigma ^2}{{\rm{I}}_{M_a}}} )^{ - 1}}{{\rm{R}}_{k,p}}$. The error covariance is ${{\rm{C}}_{k,p}} = {{\rm{R}}_{k,p}} - {\Gamma _{k,p}}$.
The computational complexity of the proposed PASCE algorithm depends on the number of iterations, sparsity, dictionary matrix size, and polarization scheme. Due to the finite sparsity, which is significantly smaller than the size of dictionary matrix, the computational complexity of the algorithm is ${\rm O}(2\mathbf{S}M{N^2})$. Approximately ${\rm O}(\mathbf{S}M{N^2})$ for the OMP algorithm.

\section{Precoding scheme design}
\subsection{Uplink Transmission}
In the uplink communication, the TAU of the train must convey the data from all on-board tasks to the BS. To mitigate interference during peak multi-user and multi-service demand, MMSE precoding can be employed for the uplink transmission. The TAU follows OTFS modulation, yields the transmission vector  ${\rm{x}}_{k,m}^{{\rm{ul}}}$. This vector is then transmitted through the wireless channel, ultimately arriving at the AP. The AP subsequently transforms the time-domain signal into a DD signal, resulting in the reception vector ${\rm{y}}_{k,m}$. Subsequent processing of this vector reveals the aggregate of all user transmissions, which can be represented as follows
\vspace{-0.5em}
\begin{equation}
{\rm{y}}_{k,m}^{{\rm{ul}}} = \sum\limits_{k = 1}^{{\cal K}} {\sqrt {{P_{k,m}}} {{\rm{H}}_{k,m}}{\rm{x}}_{k,m}^{{\rm{ul}}}}  + {{\rm{w}}_{k,m}},
\end{equation}
where ${{{\rm{H}}_{k,m}}}$ is the equivalent channel matrix between the TAU and the AP, ${{\rm{w}}_{k,m}}$ is the noise. Define the uplink power matrix as ${\rm{P}}_k^{{\rm{ul}}} = diag(P_{k,V}^{{\rm{ul}}},P_{k,H}^{{\rm{ul}}})$. The precoding matrix for the MMSE merge vector is ${\rm{v}}_{k,p}^{{\rm{}}} = {(\sum\limits_{k = 1}^{{{\cal K}}} {{{{\rm{\hat H}}}_{k,p}}} {\rm{P}}_k^{{\rm{ul}}}{\rm{\hat H}}_{k,p}^H + \sigma _{{\rm{ul}}}^2{\rm{I}})^{ - 1}}{{{\rm{\hat h}}}_{k,p}}$, where $p \in \{ V,H\}$. The received signal is given as
\vspace{-0.5em}
\begin{equation}
\begin{split}
{\rm{v}}_{k,p}^{{\rm{}}}y_{k,p}^{{\rm{ul}}} = & \underbrace {\sqrt {P_{k,p}^{{\rm{ul}}}}\mathbb{E} \{ {{({\rm{v}}_{k,p}^{{\rm{}}})}^H}{{\rm{h}}_{k,p}}\} {x_{k,p}}}_{{\rm{Desired \quad signal}}} \\       
& + \underbrace {\sqrt {P_{k,p}^{{\rm{ul}}}} ({{({\rm{v}}_{k,p}^{{\rm{}}})}^H}{{\rm{h}}_{k,p}} - \mathbb{E}\{ {{({\rm{v}}_{k,p}^{{\rm{}}})}^H}{{\rm{h}}_{k,p}}\} ){x_{k,p}}}_{{\rm{Estimation \quad error \quad interference}}} \\ 
& + \underbrace {\sum\limits_{l \ne k} {\sum\limits_{q \in \{ V,H\} } {\sqrt {P_{l,q}^{{\rm{ul}}}} {{({\rm{v}}_{k,p}^{{\rm{}}})}^H}{{\rm{h}}_{l,q}}{x_{k,p}}} } }_{{\rm{Multi-use \quad interference \quad (MUI)}}} \\
& + \underbrace {\sum\limits_{q \ne p} {\sqrt {P_{k,q}^{{\rm{ul}}}} {{({\rm{v}}_{k,p}^{{\rm{}}})}^H}{{\rm{h}}_{k,q}}{x_{k,p}}} }_{{\rm{XPC}}} + \underbrace {n_{k,p}^{{\rm{ul}}}}_{{\rm{noise}}}.
\end{split}
\end{equation}

The signal-to-interference and noise ratio (SINR) for the $k$th user equipment (UE) is
\begin{equation}
{\rm{SINR}}_{k,p}^{{\rm{ul}}} = \frac{{P_{k,p}^{{\rm{ul}}}|\mathbb{E}\{ {{({\rm{v}}_{k,p}^{{\rm{}}})}^H}{{\rm{h}}_{k,p}}\} {|^2}}}{{I_{{\rm{est,}}k,p}^{{\rm{ul}}} + I_{{\rm{MUI,}}k,p}^{{\rm{ul}}} + I_{{\rm{XPC,}}k,p}^{{\rm{ul}}} + \sigma _{{\rm{ul}}}^2\mathbb{E}\{ ||{\rm{v}}_{k,p}^{{\rm{}}}|{|^2}\} }},
\end{equation}
where $I_{{\rm{est,}}k,p}^{{\rm{ul}}} = P_{k,p}^{{\rm{ul}}}\mathbb{E}\{ |{({\rm{v}}_{k,p}^{{\rm{}}})^H}{{\rm{h}}_{k,p}} - \mathbb{E}\{ {({\rm{v}}_{k,p}^{{\rm{}}})^H}{{\rm{h}}_{k,p}}\} {|^2}\}$ denotes channel estimation error interference, $I_{{\rm{MUI,}}k,p}^{{\rm{ul}}} = \sum\limits_{l \ne k} {\sum\limits_{q \in \{ V,H\} } {P_{l,q}^{{\rm{ul}}}\mathbb{E}\{ |{{({\rm{v}}_{k,p}^{{\rm{}}})}^H}{{\rm{h}}_{l,q}}{|^2}\} } }$ indicates multi-user interference, and $I_{{\rm{XPC,}}k,p}^{{\rm{ul}}} = \sum\limits_{q \ne p} {P_{k,q}^{{\rm{ul}}}\mathbb{E}\{ |{{({\rm{v}}_{k,p}^{{\rm{}}})}^H}{{\rm{h}}_{k,q}}{|^2}\} }$
indicates cross-polarization interference,caused by signal leakage from different polarization modes, strongly correlated with the parameters of XPD and XPC. \par
By using the MMSE precoding scheme, the achievable SE of user $k$ can be expressed as 
\vspace{-0.5em}
\begin{equation}
R_k^{{\rm{ul}}} = (1 - \frac{{{\tau _p}}}{{{\tau _c}}})\sum\limits_{p \in \{ V,H\} } {{{\log }_2}(1 + {\rm{SINR}}_{k,p}^{{\rm{ul}}})}.
\end{equation}\par
The achievable SE in Eq. (37) can be computed in closed form as given in Eq. (38), shown at the bottom of the page.
\begin{figure*}[hb]
	\centering
	\vspace*{-0.5em}
	\hrulefill
	\vspace*{-0.5em}
	\begin{align}
\begin{array}{l}
R_k^{{\rm{ul}}} = (1 - \frac{{{\tau _p}}}{{{\tau _c}}})\sum\limits_{p \in \{ V,H\} } {{{\log }_2}(1 + \frac{{P _{k,p}^{{\rm{ul}}}{{[(1/{M}_{a}){\rm{tr(}}{\Gamma _{k,p}}{\rm{)}}]}^2}}}{{P _{k,p}^{{\rm{ul}}}\sum\nolimits_{l \ne k} {\frac{1}{{M}_{a}}{\rm{tr(}}{{\rm{R}}_{k,p}}{\Gamma _{k,p}}{{\rm{R}}_{k,p}}{\rm{)}}/{{[1 + \frac{1}{{M}_{a}}{\rm{tr(}}{\Gamma _{k,p}}{\rm{)}}]}^2}}  + \frac{{P_{k,p}^{{\rm{ul}}}}}{{M}_{a}}{\rm{tr(}}{{\rm{C}}_{k,p}}{\rm{)}} + {\sigma ^2}}})}
\end{array}
	\end{align}
\end{figure*} \par

\subsection{Downlink Transmission}
The downlink signal gains advantages from channel estimation. Given that the base station must handle a substantial volume of uplink signals and has finite computational resources, MR precoding is favored when downlink interference is minimal. Additionally, MMSE and ZF precoding methods are also taken into account.\par
In the downlink, the AP transmits data symbol ${{{\rm{x}}_{k,p}}} \in {^{MN \times 1}}$ to all $\cal K$ users aboard the train via a precoded matrix ${\rm{W}}_k^{{\rm{MR}}}$. The MR precoding matrix can be expressed as
\vspace{-0.5em}
\begin{equation}
{\rm{W}}_{k,p}^{{\rm{MR}}} = \frac{{{{{\rm{\hat h}}}_{k,p}}}}{{\sqrt {\mathbb{E}||{{{\rm{\hat h}}}_{k,p}}|{|^2} + \sum\limits_{l \ne k} {\sum\limits_{q \in \{ V,H\} } \mathbb{E}{\{ ||{{{\rm{\hat h}}}_{l,q}}|{|^2}\} } } } }},
\end{equation}
where $p \in \{ V,H\} $ and ${{{{\rm{\hat h}}}_{k,p}}}$ denotes the estimated channel. When normalizing for power, the expression can be further expressed as
\vspace{-0.5em}
\begin{equation}
{\rm{W}}_k^{{\rm{MR}}} = \left[ {\begin{array}{*{20}{c}}
{{\rm{W}}_{k,V}^{{\rm{MR}}}}&{{\rm{W}}_{k,H}^{{\rm{MR}}}}
\end{array}} \right]\left[ {\begin{array}{*{20}{c}}
{\sqrt {P_{k,V}^{{\rm{dl}}}} }&0\\
0&{\sqrt {P_{k,H}^{{\rm{dl}}}} }
\end{array}} \right].
\end{equation}
The downlink transmission signal can be expressed as follows
\vspace{-0.5em}
\begin{equation}
{\rm{x}}_l^{{\rm{dl}}} = \sum\limits_{k = 1}^{\cal K} {\sum\limits_{p \in \{ V,H\} } {{\rm{W}}_{k,p}^{{\rm{MR}}}{{\rm{x}}_{k,p}}} }
\end{equation}

By following propagation through the V2G wireless channel, the received signal can be represented as Eq. (42). The noise term obeys the distribution $n_{k,p}^{{\rm{dl}}}\sim{\cal C}{\cal N}\left( {{\bf{0}},\sigma _{{\rm{dl}}}^2{{\rm{I}}_2}} \right)$.

\vspace{-0.5em}
\begin{equation}
\begin{split}
{\rm{W}}_{k,p}^{{\rm{MR}}}y_{k,p}^{{\rm{dl}}} = & \underbrace {\sqrt {P_{k,p}^{{\rm{dl}}}} \mathbb{E}\{ {\rm{h}}_{k,p}^H{\rm{W}}_{k,p}^{{\rm{MR}}}\} {x_{k,p}}}_{{\rm{Desired \quad signal}}} \\
& + \underbrace {\sqrt {P_{k,p}^{{\rm{dl}}}} ({{\rm{h}}_{k,p}}{\rm{W}}_{k,p}^{{\rm{MR}}} - \mathbb{E}\{ {\rm{h}}_{k,p}^H{\rm{W}}_{k,p}^{{\rm{MR}}}\} ){x_{k,p}}}_{{\rm{Estimation \quad error \quad interference}}} \\
& \underbrace { + \sum\limits_{l \ne k} {\sum\limits_{q \in \{ V,H\} } {\sqrt {P_{l,q}^{{\rm{dl}}}} {\rm{h}}_{l,q}^H{\rm{W}}_{l,q}^{{\rm{MR}}}{x_{l,q}}} } }_{{\rm{MUI}}} + \underbrace {n_{k,p}^{{\rm{dl}}}}_{{\rm{noise}}}.
\end{split}
\end{equation}

The achievable SE of the downlink can be expressed as
\vspace{-0.5em}
\begin{equation}
R_k^{{\rm{dl}}} = (1 - \frac{{{\tau _p}}}{{{\tau _c}}})\sum\limits_{p \in \{ V,H\} } {{{\log }_2}(1 + {\rm{SINR}}_{k,p}^{{\rm{dl}}})}.
\end{equation}\par
By explicitly separating the interfering components, the downlink ${\rm{SINR}}_{k,p}^{{\rm{dl}}}$ can be defined as Eq. (44) and the closed form as given in Eq. (45), which are shown at the bottom of the next page.

\begin{figure*}[hb]
	\centering
	\vspace*{-0.5em}
	\hrulefill
	\vspace*{-0.5em}
	\begin{align}
\begin{array}{l}
{\rm{SINR}}_{k,p}^{{\rm{dl}}} = \dfrac{{P_{k,p}^{{\rm{dl}}}|\mathbb{E}\{ {\rm{h}}_{k,p}^H{\rm{W}}_{k,p}^{{\rm{MR}}}\} {|^2}}}{{P_{k,p}^{{\rm{dl}}}\mathbb{E}\{ |{{\rm{h}}_{k,p}}{\rm{W}}_{k,p}^{{\rm{MR}}} -\mathbb{E} \{ {\rm{h}}_{k,p}^H{\rm{W}}_{k,p}^{{\rm{MR}}}\} {|^2}\}  + \sum\limits_{l \ne k} {\sum\limits_{q \in \{ V,H\} } {P_{l,q}^{{\rm{dl}}}\mathbb{E}\{ |{\rm{h}}_{l,q}^H{\rm{W}}_{l,q}^{{\rm{MR}}}{|^2}\} } }  + \sigma _{{\rm{dl}}}^2\mathbb{E}\{ ||{\rm{W}}_{k,p}^{{\rm{MR}}}|{|^2}\} }}.
\end{array}
	\end{align}
\end{figure*} 

\begin{figure*}[hb]
	\centering
	\vspace*{-0.5em}
	\hrulefill
	\vspace*{-0.5em}
	\begin{align}
\begin{array}{l}
R_k^{{\rm{dl}}} = (1 - \frac{{{\tau _p}}}{{{\tau _c}}})\sum\limits_{p \in \{ V,H\} } {{{\log }_2}(1 + \frac{{P _{k,p}^{{\rm{ul}}}{{[(1/{M}_{a}){\rm{tr(}}{\Gamma _{k,p}}{\rm{)}}]}^2}}}{{P _{k,p}^{{\rm{ul}}}\sum\nolimits_{l = k} {{\rm{tr(}}{{\rm{R}}_{k,p}}{\Gamma _{k,p}}{{\rm{R}}_{k,p}}{\rm{)}}/{\rm{tr(}}{\Gamma _{k,p}}{\rm{)}}}  + \frac{1}{{M}_{a}}{\rm{tr(}}{{\rm{C}}_{k,p}}{\rm{)}} + {\sigma ^2}}})}
\end{array}
	\end{align}
\end{figure*} 
\par
In order to improve the SE of the system, it is necessary to design the corresponding interference cancellation algorithms. This will be introduced in the next subsection.

\subsection{Design of Polarized-Aware Dynamic Interference Cancellation Algorithm}

\renewcommand{\algorithmicrequire}{\textbf{Input:}}
\renewcommand{\algorithmicensure}{\textbf{Initialize:}}

\begin{algorithm}[t]
\caption{Polarized-Aware Dynamic Interference Cancellation Algorithm with Dual-Polarized precoding}
\begin{algorithmic}[1]
\REQUIRE Estimated channel matrix $\hat{\mathbf{H}} \in \mathbb{C}^{N_r \times 2K}$, Recovered polarized channel vectors $\hat{\mathbf{h}}_s^{\mathcal{P}}$, noise power $\sigma^2$, power allocation matrix $\mathbf{P} = \text{diag}(P_{1,V}, P_{1,H}, \dots, P_{K,V}, P_{K,H})$.

\ENSURE Received signal $\mathbf{y}_{\text{r}} = \mathbf{y}$, detected symbols $\hat{\mathbf{x}} = \emptyset$, 
ordered user set $\mathcal{O} = \text{sort}(\|\hat{\mathbf{h}}_{k,V}\|^2 + \|\hat{\mathbf{h}}_{k,H}\|^2)$, 
active user set $\mathcal{S} = \mathcal{O}$.

    \FOR{$\text{user}$ $k$ = $1$ to $K$}
        \STATE Select dominant user: $k^* = \arg\max_{k \in \mathcal{S}} \|\hat{\mathbf{h}}_{k,V}\|^2 + \|\hat{\mathbf{h}}_{k,H}\|^2$;
        
        \FOR{$p \in \{V, H\}$}
            \STATE Compute interference covariance using Eq. (47);
            \STATE Compute MMSE precoding vector using ${\rm{v}}_{k,p}$;
            \STATE Compute MR precoding vector using Eq. (39);
            \STATE Detect symbol: $\hat{x}_{k^*,p} = \mathbf{W}_{k^*,p}^H \mathbf{y}_{\text{r},p}$;
        \ENDFOR
        
        \STATE Update detected symbols: $\hat{\mathbf{x}} \gets \hat{\mathbf{x}} \cup \{\hat{x}_{k^*,V}, \hat{x}_{k^*,H}\}$;
        
        \STATE Polarized interference reconstruction:
        \vspace{-0.5em}
        \[
        \mathbf{y}_{\text{r}} \gets \mathbf{y}_{\text{r}} - \sum_{p\in\{V,H\}} \sqrt{P_{k^*,p}}\hat{\mathbf{h}}_{k^*,p}\hat{x}_{k^*,p};
        \]
        
        \STATE Dynamic set update: $\mathcal{S} \gets \mathcal{S} \setminus \{k^*\}$;
    \ENDFOR
    
    \STATE Calculate the SE of the MMSE precoding using Eq. (38).
    \STATE Calculate the SE of the MR precoding using Eq. (45).

    \STATE \textbf{Output:} $\hat{\mathbf{x}}$, $R_{\text{sum}}$.
\end{algorithmic}
\end{algorithm}

We sort the user channel gains in descending order based on the sum of the squares of the norms of the channel vectors for the two polarization directions. Then, we detect the strongest user signal from the residual signal and apply filtering to the V and H polarization directions separately. After signal detection, we use the estimated channel and detected symbols and subtract the detected symbols from the residual signal. Different interference cancellation methods need to be considered for different precoding schemes, as shown in Algorithm 2.\par

For polarization direction $p \in \{ V,H\} $, define the active user set as $\mathcal{S}$, define the residual interference channel matrix, i.e. the channel vectors of users other than $k^*$ among the remaining users in that polarization direction are \par
\vspace{-0.5em}
\begin{equation}
\hat{\mathbf{H}}_{S\setminus \{k^*\},p} = [ \hat{\mathbf{h}}_{l,p} ]_{l \in S, l \neq k^*}.
\end{equation} \par
Then, the interference plus noise covariance matrix is
\vspace{-0.5em}
\begin{equation}
{\bf{R}}_I^p = {P_{k,p}}{\widehat {\bf{H}}_{S \setminus \{ {k^*}\} ,p}}\widehat {\bf{H}}_{S \setminus \{ {k^*}\} ,p}^H + \sigma _n^2{\bf{I}}.
\end{equation}

For different polarization modes and different user channels, the polarized-aware dynamic interference cancellation (PADIC) algorithm is used to eliminate interference one by one according to different precoding schemes, thus obtaining the optimal SE. The computational complexity of the proposed PADIC algorithm depends on the number of users, the precoding scheme, and the number of antennas. The overall complexity of the PADIC algorithm is ${\rm O}({\cal K}*2*({\cal K}{{{M}_{a}}^2} + {{{M}_{a}}^3}))$. When the number of users $\cal K$ is much smaller than the number of antennas ${{M}_{a}}$, the complexity is approximately ${\rm O}({\cal K}{{{M}_{a}}^3}))$. The complexity of the SIC algorithm is ${\rm O}({\cal K}^2{{{M}_{a}}^2} + {{\cal K}{{M}_{a}}^3})$. The complexity of PADIC is comparable to the SIC algorithm, but it is more suitable for dual-polarized systems.

\section{Simulation Analysis And Results}

In this paper, the URT tunnel scene is selected, in which the tunnel cross section is arched. The height of the tunnel is 5 m and the width at the bottom is 3.4 m \cite{ref28}. The height of the TAU is 3.8 m, and the height of the AP is defined at 4.2 m. The distance between each AP is 10 meters. The carrier frequency is selected as 1.8 GHz for LTE-M, with a bandwidth of 20 MHz \cite{ref2}. The vehicle's operating speed is defined as 120 km/h. $M = 128$, $N = 16$. The large-scale attenuation coefficient is defined as ${\beta _{m,n}}(t)$, and can be expressed as
\vspace{-0.5em}
\begin{equation}
{\beta _{m,n}}(t) = {10^{\frac{{P{L_{m,n}}(t)}}{{10}}}}{10^{\frac{{{\sigma _{SF}}{\mu _{m,n}}(t)}}{{10}}}},
\end{equation}
where ${P{L_{m,n}}(t)}$ denotes the path loss between the TAU and the AP at time $t$. ${10^{\frac{{{\sigma _{SF}}{\mu _{m,n}}(t)}}{{10}}}}$ represents the shadow fading with standard deviation ${\sigma _{SF}}$ and ${\mu _{m,n}}(t)$ is the fading coefficient, it follows the Gaussian distribution ${\cal N} \sim \left( {0,1} \right)$. \par

In Fig. 3, the channel model for LCX also adopts the GBSM theory, with an LCX slot spacing of $3\lambda $. Under high signal-to-noise ratios (SNR) condition, dual-polarized MIMO can enhance spectral efficiency by 78.8\% to 104.9\% relative to LCX. For dual-polarized MIMO systems, doubling the number of antennas nearly doubles the system's SE. However, a significant XPC can severely impact the SE of dual-polarized systems even with the same number of antennas. Therefore, it is essential to employ precoding schemes to suppress XPC effects and enhance the SE of the system.\par

Fig. 4 illustrates a comparative analysis of channel estimation efficacy across various algorithms, including the sparse factor suppression (SFS) method, the OMP algorithm, and the PASCE algorithm proposed in this paper. The SFS algorithm is used to obtain the amplitude of the received signal by estimation, determine an appropriate suppression factor, and suppress a large number of low-amplitude signals. The result of this process is the extraction of sparse but valuable signals. As can be seen in the figure, the PASCE algorithm proposed in this paper is superior to the comparison algorithm. When the adaptive adjustment sparse step size is defined as $1$, relatively accurate channel estimation can be performed.\par
\begin{figure}[!ht]
\centering
\includegraphics[width=3.2in]{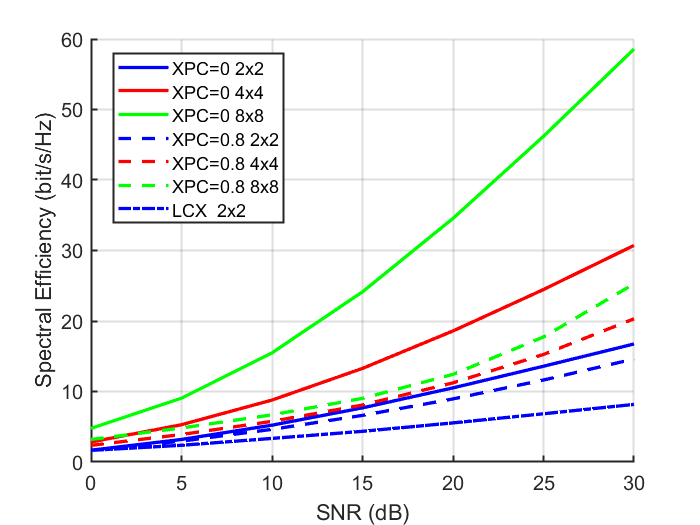}
\caption{Comparison of system SE under different MIMO configurations and XPC values.}
\label{fig_3}
\end{figure}

\begin{figure}[!ht]
\centering
\includegraphics[width=3.2in]{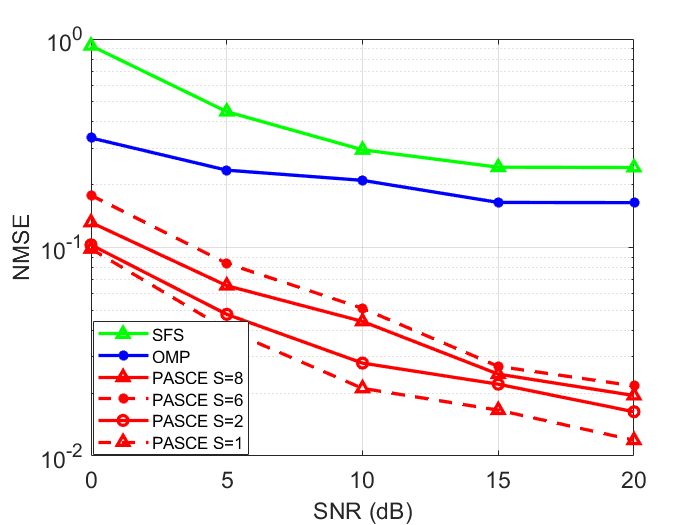}
\caption{Comparison of NMSE performance of different algorithms for channel estimation.}
\label{fig_4}
\end{figure}

Fig. 5 shows that when different MIMO configurations and precoding schemes. The scheme that without precoding exhibits the lowest SE when the number of antenna configuration is 2×2, while the SE of the precoding schemes using MMSE-PADIC, MR-PADIC and ZF-PADIC gradually increases. When the number of antenna configuration is 4×4, without precoding scheme still exhibits the lowest SE. Overall, the MMSE-PADIC precoding scheme is optimal. In low SNR condition, the performance of the ZF-PADIC precoding scheme is almost identical to that of the non-precoding scheme. The MR-PADIC precoding scheme performs well under low SNR, but its performance deteriorates slightly compared to MMSE-PADIC under high SNR.

\begin{figure}[!ht]
\centering
\includegraphics[width=3.2in]{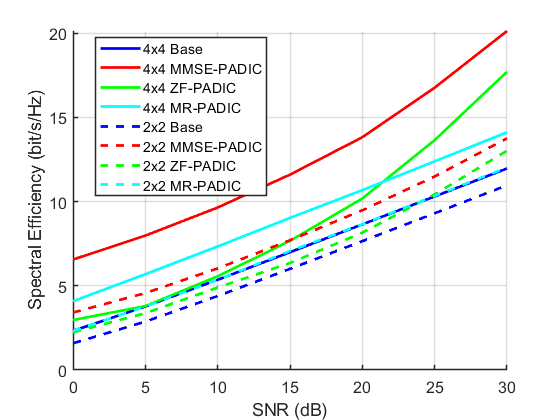}
\caption{Comparison of system SE performance for different precoding schemes.}
\label{fig_5}
\end{figure}

Fig. 6 shows the coverage of the single-polarized antenna channels, as well as 3GPP's Extended Vehicular A (EVA), Extended Typical Urban (ETU), in a dual-polarized MIMO channel, a curved tunnel environment and LCX channel. This representation covers most of the operational scenarios encountered in URT, either underground or overhead. As shown in the figure, when using LCX for signal coverage, the SE of multi-user systems is relatively low. Using dual-polarized MIMO antenna can provide optimal downlink SE, ensuring the wireless communication needs of URT for multiple tasks and users. In curved tunnel scenarios, the curvature of the tunnel affects signal strength, resulting in reduced transmission rates in multi-user scenarios. When using a single-polarized antenna, the SE of the system is relatively lower compared to the dual-polarized antenna system. However, as the number of users increases, the SE of both systems tends to converge. The SE of the multi-user system under the EVA and ETU channels differs only slightly.

\begin{figure}[!t]
\centering
\includegraphics[width=3.2in]{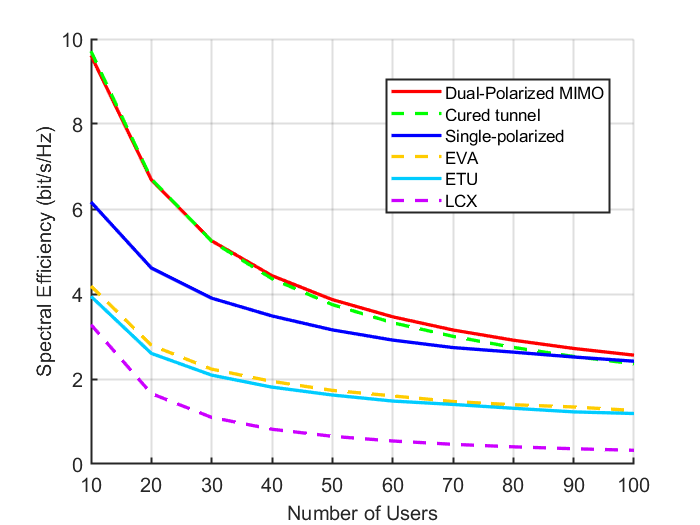}
\caption{Comparison of the number of users and system SE under different channels in the downlink.}
\label{fig_6}
\end{figure}

 In Fig. 7, as the ${{\rho _{{t_k}}}}$ and ${{\rho _{{r_k}}}}$ values increase, the non-diagonal elements of the XPC channel matrix experience an increase in enhancement, resulting in notable correlation within the channel polarization direction. This effect causes the eigenvalue distribution to become more dispersed, degrading the channel's numerical stability. The SE is increased by 21.42\%, 27.25\% and 6.9\% for the MMSE-PADIC, ZF-PADIC, and MR-PADIC precoding schemes, compared to when the XPC drops from 0.8 to 0.1, respectively. Certainly, employing precoding scheme is higher than the baseline scheme, regardless of the value of XPC.

\begin{figure}[!t]
\centering
\includegraphics[width=3.2in]{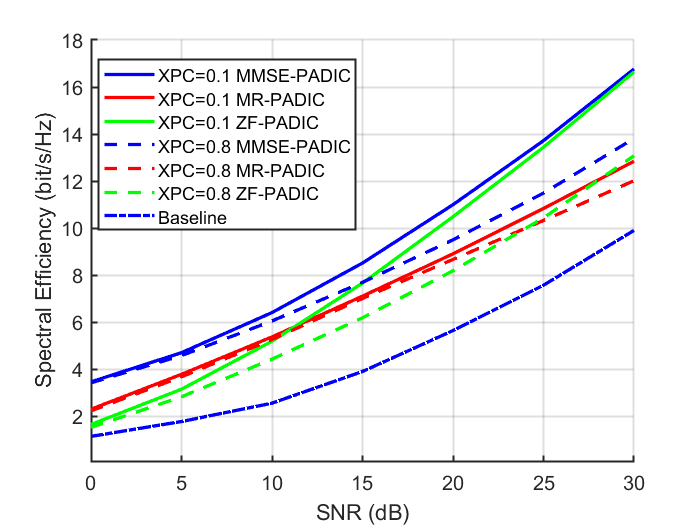}
\caption{System SE under different precoding schemes at different XPC values.}
\label{fig_7}
\end{figure}

Fig. 8 illustrates the comparison between theoretical and measured values for the different precoding schemes proposed in this paper. The measured data and channel model employ a single-slope model for large-scale fading and approximate the small-scale fading envelope as a Nakagami distribution \cite{ref29}. It reveals that the trends of the theoretical and measured values align closely. The MMSE-PADIC, MR-PADIC, and ZF-PADIC precoding schemes yield values 11.63\%, 29.25\%, and 30.8\% higher than the measured values, respectively, demonstrating the superiority of the proposed algorithms. The black dashed line in the figure represents the convergence performance of the algorithm without SIC interference cancellation, demonstrating the excellent convergence characteristics of the proposed algorithm.

\begin{figure}[!t]
\centering
\includegraphics[width=3.2in]{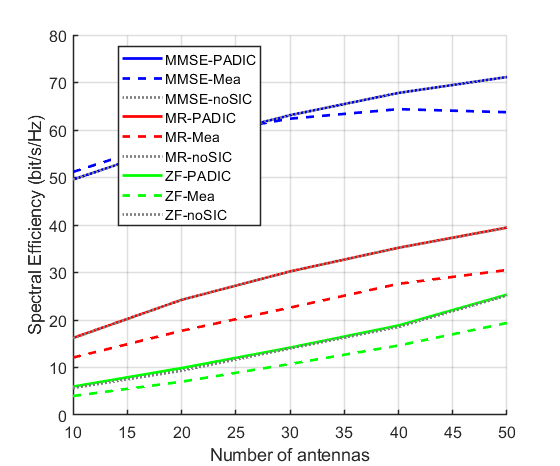}
\caption{Comparison of system antenna numbers and SE for different precoding schemes.}
\label{fig_8}
\end{figure}

\section*{Conclusion}
In this paper, we propose an innovative dual-polarized distributed MIMO system architecture to improve the communication rates for V2G wireless communication in URT tunnel scenarios. To accurately characterize the propagation mechanism of signals within tunnels, we employ GBSM theory for channel modeling. This approach considers both the placement of the transmitting and receiving devices within the tunnel and the dynamics of the movement of the train. We propose the PASCE algorithm for effective channel estimation. Based on the estimated channel, we derive closed-form solutions for MMSE and MR precoding schemes, as well as uplink and downlink SE. We employ the PADIC algorithm to eliminate XPC and MUI, thus improving the system's SE.\par
We conducted simulation analyses for different channel models, XPC effects, and antenna configurations. Compared with the LCX channel based on GBSM, the dual-polarized MIMO channel demonstrates superiority. The PADIC algorithm proposed in this paper maintains good system performance even under high XPC effects. The simulated values from the algorithm closely match the measured values.

\vfill

\end{document}